\documentclass{article}

\usepackage{fullpage}
\usepackage{natbib}
\usepackage{authblk}
\usepackage[english]{babel}
\usepackage[utf8]{inputenc}
\usepackage{graphicx}
\usepackage{amsmath}
\usepackage{xcolor}
\usepackage[noabbrev]{cleveref}
\usepackage[labelformat=parens,subrefformat=parens]{subfig}
\usepackage{siunitx}
\usepackage{caption}
\usepackage{amssymb}
\usepackage{xcolor}
\usepackage{layouts}
\usepackage{mathtools} 
\usepackage{tikz}
\usepackage{placeins}
\usepackage{tabularx}
\newcolumntype{L}[1]{>{\raggedright\arraybackslash}p{#1}}
\newcolumntype{C}[1]{>{\centering\arraybackslash}p{#1}}
\newcolumntype{R}[1]{>{\raggedleft\arraybackslash}p{#1}}
\usepackage{booktabs}
\usepackage{supertabular}
\usepackage{tabularx}
\usepackage{booktabs}
\usepackage{array}
\usepackage{bm}

\newcommand{\LP}{$\lambda^+$}
\newcommand{\TP}{$T^+$}
\newcommand{\AP}{$A^+$}
\newcommand\crule[3][black]{\textcolor{#1}{\rule{#2}{#3}}}

\newcommand{\fig}[1]{\crule[yellow]{1em}{1em}}

\begin{document}

\title{Actuation response model from sparse data for wall turbulence drag reduction}

\author[1]{Daniel Fernex\footnote{\textit{d.fernex@tu-braunschweig.de}}}
\author[1]{Richard Semaan}
\author[2]{Marian Albers}
\author[2]{Pascal S.\ Meysonnat}
\author[2,3]{Wolfgang Schröder}
\author[4,5,1,6]{Bernd R.\ Noack}
\affil[1]{Institut für Strömungsmechanik, Technische Universität Braunschweig, Hermann-Blenk-Str. 37, 38108 Braunschweig, Germany}
\affil[2]{Institute of Aerodynamics, RWTH Aachen University, Wüllnerstr.  5a, 52062 Aachen, Germany}
\affil[3]{JARA – High-Performance Computing, Forschungszentrum Jülich, 52425 Jülich, Germany}
\affil[4]{Institut für Strömungsmechanik und Technische Akustik (ISTA), Technische Universität Berlin, Müller-Breslau-Str. 8, 10623 Berlin, Germany}
\affil[5]{LIMSI, CNRS, Université Paris-Saclay, Bât 507, rue du Belvédère, Campus Universitaire, F-91403 Orsay, France}
\affil[6]{Institute for Turbulence-Noise-Vibration Interaction and Control, Harbin Institute of Technology, Shenzhen Campus, China}

\date{\today}

\maketitle

\begin{abstract}
  We compute, model, and predict drag reduction of an actuated
  turbulent boundary layer at a momentum thickness based Reynolds number of $Re_{\theta}=1000$.  The actuation is performed using spanwise
  traveling transversal surface waves parameterized by wavelength,
  amplitude, and period. The drag reduction for the set of actuation
  parameters is modeled using 71 large-eddy simulations (LES).  This
  drag model allows to extrapolate outside the actuation domain
  for larger wavelengths and amplitudes.
  The modeling novelty is based on combining support vector regression for interpolation, a
  parameterized ridgeline leading out of the data domain, scaling from
  \cite{tomiyama_direct_2013}, and a discovered self-similar structure of the
  actuation effect.  The model yields high prediction accuracy outside the
  training data range.
\end{abstract}

\section{Introduction}

The skin friction associated with a turbulent boundary layer
constitutes about \SI{50}{\percent} of the total drag of an airplane.
Owing to its importance, passive or
active skin-friction reduction means have widely been investigated \citep{Fan2016book}.
Promising strategies include riblets
\citep{walsh_optimization_1984}, compliant surfaces \citep{Luhar2016jot}, spanwise wall oscillations and
similar variations
\citep{jung_suppression_1992,quadrio_streamwise-travelling_2009}, and
spanwise traveling waves of spanwise forcing
\citep{du_suppressing_2000} or wall-normal deflection
\citep{Klumpp2011ftc,Albers2019,Albers2019Arxiv}.
To determine the optimal actuation settings, the parameter
space is typically scanned by performing a large number of numerical
simulations, which is very costly and sometimes untractable,
e.g., for high Reynolds number or large actuation wavelength.  This
reliance on numerical simulations is partially due to the fact that
experiments for many of these actuation concepts are either currently
unrealizable or are limited by design to a small actuation range. In
this study, a methodology is developed to model sparse flow response
data to spanwise traveling surface waves using a machine learning
regression algorithm for interpolation and a ridgeline modeling for
extrapolation and optimization, which reduces the necessity for a
large parametric study.  Investigations on the boundary layer response
sensitivities show a self-similar response behavior starting at a
certain wavelength.

Actuation employing a spanwise pressure gradient has been shown to
attenuate the boundary layer low-speed streaks and reduce turbulence
production.  This principle was first put into practice by a spanwise
wall oscillation for turbulent channel flows
\citep{jung_suppression_1992,touber_near-wall_2012} and for turbulent
boundary layers \citep{lardeau_streamwise_2013}.  The actuation
generates a thin spanwise Stokes layer and reduces the wall-shear
stress.  A more efficient variant of the spanwise actuation is the
streamwise traveling wave of spanwise forcing
\citep{quadrio_streamwise-travelling_2009}, where a maximum drag
reduction of \SI{48}{\percent} can be achieved for turbulent channel
flows.  Using the same actuation technique
\cite{gatti_reynolds-number_2016} performed a comprehensive study of
4020 direct numerical simulations of a channel flow with varying
wavenumber, amplitude, frequency, and Reynolds number.  Such a large
parameter study using high-fidelity simulations is unusual and
computationally very expensive.
The results showed that, for channel flows, the
drag reduction at higher Reynolds numbers can
be estimated using the vertical shift of the logarithmic velocity
profile.

Another actuation variant employs the transversal traveling wave.  A
first implementation was conducted for a channel flow by
\cite{du_suppressing_2000}, where the wave effect was generated with a
Lorentz force.  To enable real-life applications, more recent studies
proposed a similar traveling wave effect by means of surface
deformation.  This approach has been experimentally tested for a
turbulent boundary layer by \cite{li_parametric_2018}, where the
surface was deflected using electromagnetic actuators.  They achieved
a drag reduction of \SI{4.5}{\percent}.  Spanwise traveling
transversal surface waves have also been numerically simulated for a
turbulent boundary layer over a flat plate \citep{Albers2019Arxiv}, with a maximum
drag reduction of \SI{26}{\percent}, and
over a wing section \citep{Albers2019}, where the pressure varies
in the streamwise direction.  For the wing flow, the total drag
was reduced by $7.5\,\%$ and a slight lift increase was also achieved.

Drag reduction optimization in a rich actuation space constitutes a
challenge.  In experimental setups, many degrees of freedom are fixed
by design, whereas high-fidelity high Reynolds number
simulations are able to explore a rich spectrum of actuation settings.
However, high-fidelity numerical computations are costly and thus limited to a
small number of control laws.
Surrogate models are computationally cheap models that approximate the
behavior of complex systems, based on a limited number of data. Surrogate
models are typically used for optimization
\citep{Forrester2009pas,Yondo2018pas} and for visualization and design space
analysis \citep{Holden2004phd}. There exists many approaches and algorithms.
The Response Surface Methodology (RSM) is one of the earliest approaches
\citep{box_experimental_1951}.  The models from RSM are often
polynomials up to second order
\citep{Sevant2000joa,madsen_response_2000,ahn_response_2001,karami_experimental_2016},
which can not represent highly non-linear or multi-modal design landscapes.
Radial basis functions (RBF) is an interpolation technique based on a weighted
sum of radial basis functions \citep{Broomhead1988}. Various types of basis
functions can be used to accommodate the response surface complexity
\citep{Forrester2009pas}. RBF surrogate models have successfully been used to
optimize groundwater remediation design \citep{Akhtar2016jgo}.
Another widely used technique  is
kriging, which is a kernel-based probabilistic approach \citep{Matheron1963eg}.
Kriging models can yield high predictive power \citep{Forrester2009pas}
and have been used, for instance, to optimize a $2D$ airfoil geometry \citep{Jeong2005joa}.

Other computationally more expensive algorithms include support vector
regression (SVR) \citep{Drucker1997anips},  and artificial neural network
(ANN), first developed by \cite{Mcculloch1943bmp}.
%
SVR is a kernel-based regression technique, which tolerates predictions errors
within an user-defined interval. More details about SVR are given in section
\ref{Sec:SVR}.
SVR has been shown to outperform RSM, kriging, and RBF for a test bed of 26 complex
engineering functions \citep{Clarke2005jmd}, and has successfully been used
to optimize railway wind barriers \citep{Huoyue2017smo}.
ANNs are non-linear regression models inspired by biological neural networks.
They have been used to accurately predict the drag reduction in oil pipelines
\citep{zabihi_artificial_2019}.

A common shortcoming of data-driven surrogate models is their rapidly
diminishing accuracy outside the training parameter range.
This limitation means strong disadvantages for the investigated boundary layer
application. Initial analyses have shown a higher drag reduction trend
leading beyond the training parameter space, where simulations become
increasingly less affordable.
%

In this study, we present a new modeling methodology capable of extrapolating drag
reduction beyond the parameter range.
The starting point is a
sparse set of 71 large-eddy simulations (LES) of a turbulent boundary layer
actuated by spanwise traveling transversal waves.
Our approach consists of two
steps: First, a surrogate model is built using SVR to interpolate the drag
reduction in the training parameter space. Then, extrapolation is enabled through the
identification of a ridgeline in the drag reduction response.
%
%
The model is used to analyze the actuation sensitivities
and to infer higher drag reduction and the corresponding
actuation settings.

The paper is structured as follows. The numerical method of the
high-fidelity simulations
and the computational setup of the flat plate undergoing transversal
spanwise traveling waves are defined in \cref{Sec:ComputationalSetup}. The
modeling approach is described in \cref{Sec:Methodology} for a simple
problem, before being applied to the actuated boundary layer data
in \cref{Sec:Results}.  Finally, conclusions are presented in
\cref{Sec:Conclusions}.

\section{Numerical setup}
\label{Sec:ComputationalSetup}
In this section,
the open-loop actuation study of wall turbulence drag reduction is recapitulated.
In particular, the investigated actuation parameters are enumerated.
Section \ref{Subsec:Configuration} describes the configuration,
while section \ref{Sec:NumericalMethod} details
the employed large-eddy simulation (LES) solver.

\subsection{Configuration}
\label{Subsec:Configuration}
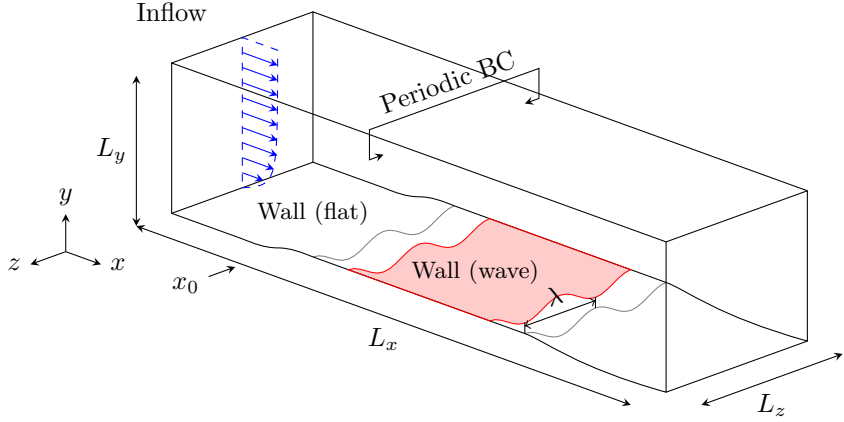
\begin{figure}
\begin{center}
  \begin{tikzpicture}[x={(0.939cm,-0.34cm)}, y={(0cm,1cm)}, z={(0.939cm,0.34cm)}]\
    \draw [->,>=stealth] (0,0,-1.5) -- (0,0,-2.0) node [left]{$z$};
    \draw [->,>=stealth] (0,0,-1.5) -- (0,0.5,-1.5) node [above]{$y$};
    \draw [->,>=stealth] (0,0,-1.5) -- (0.5,0,-1.5) node [right]{$x$};
    \draw [<->,>=stealth] (0,0,-0.5) -- (7,0,-0.5) node [pos=.5,below=1.0]{$L_x$};
    \draw [<->,>=stealth] (0,0,-0.5) -- (0,2,-0.5) node [pos=.5,left]{$L_y$};
    \draw [<->,>=stealth] (7.5,0,0) -- (7.5,0,2) node [pos=.5,below=2.0]{$L_z$};
    \draw (0,0,0) -- (1,0,0) sin (1.5,0.05,0) cos (2,0.1,0) -- (5,0.1,0) .. controls (5.3,0.1,0) and (5.7,-0.1,0) .. (7,0.0,0) -- (7,2,0) -- (7,2,0) -- (0,2,0) -- cycle;
    \draw (0,0,2) -- (1,0,2) sin (1.5,0.05,2) cos (2,0.1,2) -- (5,0.1,2) .. controls (5.3,0.1,2) and (5.7,-0.1,2) .. (7,0.0,2) -- (7,2,2) -- (7,2,2) -- (0,2,2) -- cycle;
    \draw[color=red,fill=red, fill opacity=0.2, domain=0:2, variable=\z]  (2.5,0.1,0) -- plot (2.5,{0.1*sin((1.5707+2*3.14159*\z) r)},\z)  -- (4.5,0.1,2) --  plot (4.5,{0.1*sin((1.5707+2*3.14159*\z) r)},2.0-\z)  -- (4.5,0.1,0) -- (2.5,0.1,0) -- cycle;
    \draw (7,0,0) -- (7,0,2);
    \draw (0,0,0) -- (0,0,2);
    \draw (0,2,0) -- (0,2,2);
    \draw (7,2,0) -- (7,2,2);
    \draw[opacity=0.5, variable=\z, samples at={0,0.05,...,2.05}]
    plot (5, {0.1*sin((1.5707+2*3.14159*\z) r)}, \z);
    \draw[opacity=0.5, variable=\z, domain=0:2]
    plot (2, {0.1*sin((1.5707+2*3.14159*\z) r)}, \z);
    \draw (5,0.1,0) -- (5,0.25,0);
    \draw (5,0.1,1) -- (5,0.25,1);
    \draw[<->] (5,0.2,0) -- (5,0.2,1) node [pos=.5,sloped,above] {$\lambda$};

    \node (x0) at (1.5,0,-1.3) {$x_0$};
    \draw[->,>=stealth] (x0) -- (1.5,0,-0.6);
    \node (inflow) at (-1,2,1) {Inflow};
    \node (flat) at (1, 0,1) {\small{Wall (flat)}};
    \node (flat) at (3.3, 0,1) {\small{Wall (wave)}};
    \draw[<->,>=stealth] (3,1.8,0) -- (3,1.8,-0.2) -- (3,2.2,-0.2) -- node[pos=.5,sloped,above] {Periodic BC} (3,2.2,2.2) -- (3,1.8,2.2) -- (3,1.8,2.0);
    \draw[color=blue, dashed] (0,0,1) .. controls (0.5,0.2,1) ..  (0.5,2,1) -- (0,2,1) -- (0,0,1);
    \draw[color=blue, ->,>=stealth] (0,0.2,1) -- (0.3,0.2,1);
    \draw[color=blue, ->,>=stealth] (0,0.4,1) -- (0.45,0.4,1);
    \draw[color=blue, ->,>=stealth] (0,0.6,1) -- (0.5,0.6,1);
    \draw[color=blue, ->,>=stealth] (0,0.8,1) -- (0.5,0.8,1);
    \draw[color=blue, ->,>=stealth] (0,1.0,1) -- (0.5,1.0,1);
    \draw[color=blue, ->,>=stealth] (0,1.2,1) -- (0.5,1.2,1);
    \draw[color=blue, ->,>=stealth] (0,1.4,1) -- (0.5,1.4,1);
    \draw[color=blue, ->,>=stealth] (0,1.6,1) -- (0.5,1.6,1);
    \draw[color=blue, ->,>=stealth] (0,1.8,1) -- (0.5,1.8,1);
  \end{tikzpicture}
  \caption{Overview of the physical domain of the actuated turbulent
    boundary layer flow, where $L_x, L_y,$ and $L_z$ are the domain
    dimensions in the Cartesian directions, $\lambda$ is
    the wavelength of the spanwise traveling wave, and $x_0$ marks the
    actuation onset. The shaded red surface $A_\mathrm{surf}$ marks the integration area of
    the wall-shear stress $\tau_w$.}
  \label{fig::grid}
\end{center}
\end{figure}

The fluid flow is described in a Cartesian frame of reference where
the streamwise, wall-normal, and spanwise coordinates are denoted by
$\mathbf{x} = (x,y,z)^T$ and the velocity components by $\mathbf{u} =
(u,v,w)^T$. The Mach
number is set to $M = 0.1$ such that nearly incompressible
flow is considered. An illustration of the rectangular physical domain is
shown in \cref{fig::grid}. A momentum thickness of $\theta = 1$ at
$x_0$ is achieved such that the momentum thickness based Reynolds
number is $Re_\theta = 1000$ at $x_0$. The domain length and height in the
streamwise and wall-normal direction are $L_x = 190\,\theta$ and $L_y
= 105\,\theta$. In the spanwise direction, different domain widths $L_z
\in [21.65\,\theta, 108.25\,\theta]$ are used to simulate different
actuation wavelengths.

At the domain inlet, a synthetic turbulence generation method
is applied to generate a natural turbulent boundary layer flow after a
transition length of 2-4 boundary layer thicknesses
\citep{Roidl2013}. Characteristic boundary conditions are used at the
domain exit and a no-slip wall boundary condition is enforced at
the lower domain boundary for the unactuated and actuated wall. The
wall actuation is prescribed by the space- and time-dependent
function
%
\begin{equation}
y_{\text{wall}}^+(z^+,t^+) =
A^+ \cos\left( \frac{2\pi}{\lambda^+}z^+ - \frac{2\pi}{T^+}t^+ \right)
\label{Eqn:Actuation}
\end{equation}
%
in the interval $-5 \leq x/\theta \leq 140$.
The quantities $\lambda^+$, $T^+$, and $A^+$ denote the wavelength, period, and
amplitude in inner coordinates, i.e., the parameters are scaled by the viscosity
$\nu$ and the friction velocity of the unactuated reference case
$u^n_{\tau}$.
In the area just upstream
and downstream of the wave actuation region, a spatial transition is
used from a flat plate to an actuated plate and vice versa \cite{Albers2019Arxiv}.
In total, 71 actuation configurations with wavelength $\lambda^+ \in [500,3000]$, period
$T^+ \in [20,120]$, and amplitude $A^+ \in [10,78]$ are simulated.
Two additional validation simulations are performed for $\lambda^+=5000$,
$T^+=44$, and $A^+=92$ and for $\lambda^+=5000$, $T^+=44$, and $A^+=99$.
All operating conditions and the corresponding drag reductions are listed in
appendix \ref{Sec:OCLES}.
The current setup is identical to that in \cite{Ishar2019}.  However, a
considerably larger parameter set is computed in this study, containing also larger
wavelengths.

The physical domain is discretized by a structured block-type mesh
with a resolution of $\Delta x^+ = 12.0$ in the streamwise and $\Delta
z^+ = 4.0$ in the spanwise direction. In the wall-normal direction, a
resolution of $\Delta_y^+|_{\mathrm{wall}} = 1.0$ at the wall is used
with gradual coarsening away from the wall. Depending on the
domain width, the meshes consist of $24$ to $120$ million cells.

The simulation procedure is as follows. First, the reference
simulations for all domain widths are run for $t u_\infty / \theta =
650$ convective time units. Then, the actuated simulations are
initialized by the solution from the unactuated reference case and
the temporal transition from the flat plate to the actuated wall is
initiated. When a converged state of the friction drag is obtained, statistics
are collected for $t u_\infty / \theta = 1250$ convective times.

The drag coefficient $c_d$ is computed by integrating the wall-shear
stress over the streamwise interval $50 \leq x/\theta \leq 100$ and
over the entire domain spanwise width, i.e., the colored surface
in \cref{fig::grid}
\begin{align*}
c_d &= \frac{2}{\rho_\infty u_\infty^2 A_{\mathrm{ref}}} \int_{A_\mathrm{surf}}  \tau_w \mathbf{n}\cdot\mathbf{e}_y dA~.
\end{align*}
The quantities $\mathbf{n}$, $\mathbf{e_y}$ denote the unit normal
vector of the surface and the unit vector in the $y$-direction, the
reference surface is $A_\mathrm{ref} = 1$. The
drag reduction is defined as
\begin{align*}
  J=\Delta c_d = \frac{c_{d}^{u} - c_{d}^{a}}{c_{d}^{u}}
\end{align*}
where the superscripts $u$ and $a$  refer to the unactuated
reference and actuated cases.

\subsection{Numerical method}
\label{Sec:NumericalMethod}
The actuated flat plate turbulent boundary layer flow is governed by
the unsteady compressible Navier-Stokes equations in the arbitrary
Lagrangian-Eulerian formulation for time-dependent domains. A
second-order accurate finite-volume approximation of the governing
equations is used in which the convective fluxes are computed by the
advection upstream splitting method (AUSM) and time integration is
performed via a 5-stage Runge-Kutta scheme. The smallest dissipative
scales are implicitly modeled through the numerical dissipation of the
AUSM scheme. This monotonically integrated large-eddy simulation
approach \citep{Boris1992} is capable of accurately capturing all
physics of the resolved scales \citep{Meinke2002}. Further details on
the numerical method can be found in \cite{Albers2019} and
\cite{Ishar2019}.

\section{Methodology}
\label{Sec:Methodology}
In this section,
we propose a data-driven response surface methodology
for interpolation and extrapolation.
The methodology is developed to handle the observed relative drag reduction sensitivities $J=\Delta c_D$.
We note that $J$ is \emph{positive} for \emph{reduced} drag.
Initial analyses indicate that, for the spanwise traveling wave, in every $\lambda^+$ plane, the drag reduction
$\Delta c_d$ features a single global maximum $(T_r^+,A_r^+)$ with respect to the
actuation period $T^+$ and amplitude $A^+$ of the current spanwise traveling wave type.  The curve of $(\lambda^+,T^+,A^+)$
connecting all these $\lambda^+$-dependent $\Delta c_d$ maxima is the
\emph{ridgeline}, denoted by the subscript $r$.
These optimal amplitude and period $(T_r^+,A_r^+)$ increase with $\lambda^+$ beyond the
currently simulated parameter range.  Such response behavior is challenging for
optimization and requires specially-developed tools.

\Cref{Sec:Example} introduces an analytical example
which features similar topology to the drag reduction response distribution.
The machine learning algorithm used to interpolate the response within the
parameter range is detailed in \cref{Sec:SVR}.
In \cref{Sec:Procedure}, a novel data-driven modeling approach is proposed
and exemplified for the analytical example.

\subsection{Analytical response surface}
\label{Sec:Example}
To sharpen our data-driven tools, we start with an analytic response function
$J(\bm{p})$ which behaves qualitatively similar to the drag reduction problem.
From the parameter vector $\bm{p}= (p,q,s)$, $p$ mimics the wavelength, $q$ the
period, and $s$ the amplitude.  The analytical example
%
\begin{equation}
J(p) =  \underbrace{\tanh (1+p)}_{=:G(p)} \>
     \underbrace{
     \exp \left[ -\left (1-2q/\sqrt{p} \right )^2
                 -\left (1-2s/\sqrt{p} \right )^2
         \right]}_{=:F_{p} (q,s)}
\label{Eqn:Example}
\end{equation}
is investigated in the domain
\begin{equation}
\Omega: = [0,1] \times [0,1] \times [0,1].
\label{Eqn:Domain}
\end{equation}
The function $J$ factors into one monotonously increasing term $G(p)$
and one $p$-independent monomodal term $F_{p} (q,s)$ with a single maximum in  $(q,s)$.
At a given $p$,
$J$ assumes the maximum
$J_{p} := \tanh(1+p)$
on the ridgeline
$q_r = \frac{1}{2}\sqrt{p}$,
$s_r = \frac{1}{2}\sqrt{p}$.
The ridgeline marks the maxima of $q$ and $s$
for constant $p$.
The global maximum $J_{\rm max} = \tanh (2)$
in the domain $\Omega$ is at the boundary
$\bm{p}_{\rm max} = (1,0.5,0.5)$.
%
\begin{figure}
  \begin{center}
    \includegraphics[width=0.5\linewidth]{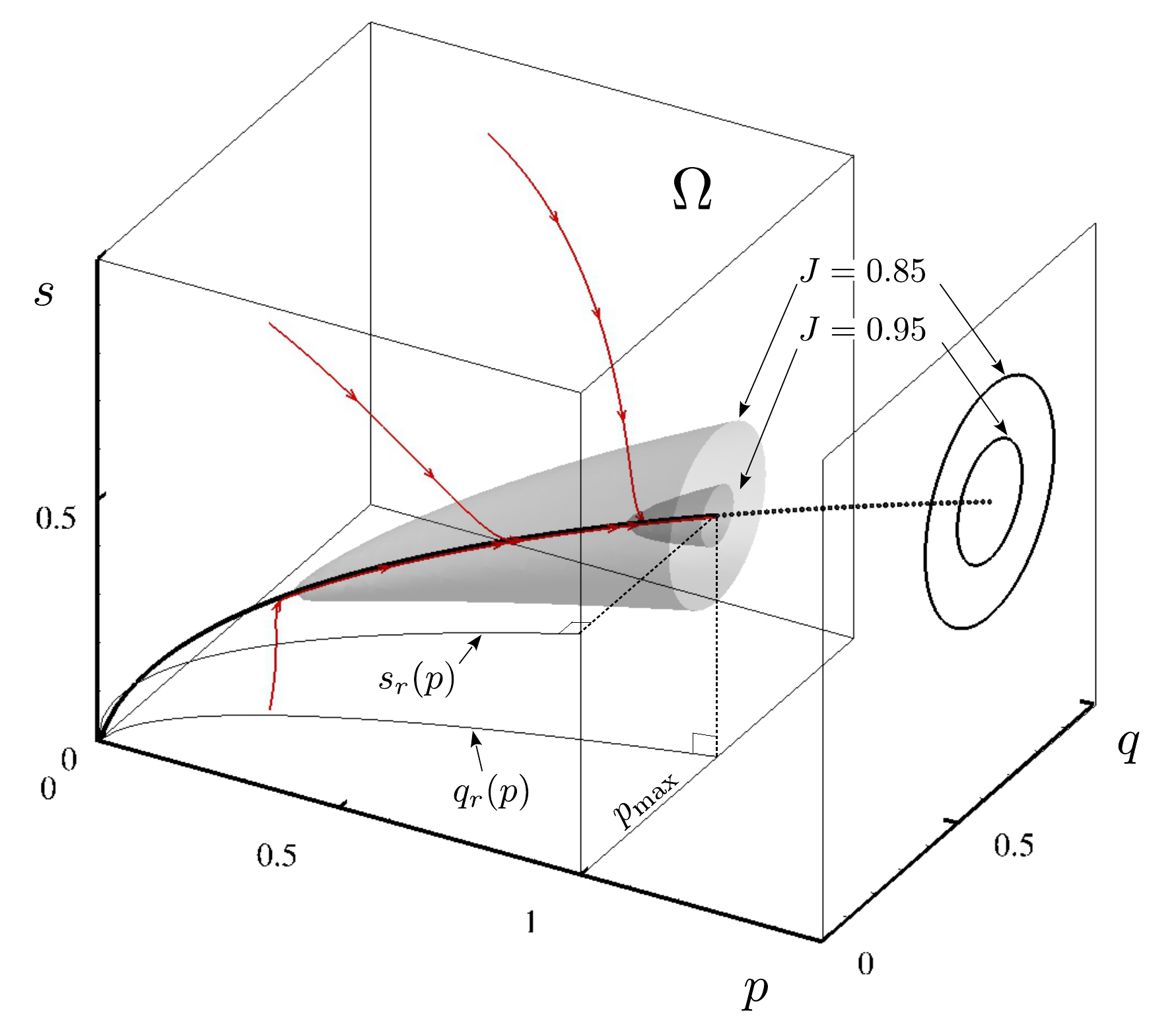}
    \end{center}
    \caption{Analytical response surface of equation \eqref{Eqn:Example}.
    Also shown is the ridgeline (black) in the interpolated (solid line) and the extrapolated (dotted line) regimes.
    The red lines denote lines of steepest ascent seeded at various domain locations.}
    \label{Fig:IsoExample}
\end{figure}
%
These trends are observed in \cref{Fig:IsoExample},
which illustrates the analytical response surface as iso-surfaces,
the ridgeline (black), and the lines of steepest ascent (red).
The lines of steepest ascent provide a direct indication to the response sensitivities
and point in the direction of the global optima.
The lines of steepest ascent are simply streamlines of the gradient field $\nabla J= (\frac{\partial J}{\partial p}, \frac{\partial J}{\partial q}, \frac{\partial J}{\partial s})$ seeded from various points.

Evidently, larger $J$ values are obtained outside
the domain $\Omega$ on the ridgeline.
Following this curve is a good extrapolation strategy
for testing new and better parameters.
The extrapolation to suboptimal parameters
outside the domain $p>p_{\rm max}$
is facilitated by the self-similar structure
of this particular response function.
The response $J$ can be parameterized by a $p$-dependent function multiplying a properly scaled $(q, s)$-dependent function
\begin{equation}
J  (p,q,s)  = J_r (p) \> F \left( q^*, s^* \right)
            = J_r (p) \> F \left( \frac{q}{q_r}, \frac{s}{s_r} \right)
\label{Eqn:Toy:Response}
\end{equation}
where
%
\begin{subequations}
\begin{eqnarray}
J_r (p) &=& \tanh(1+p),  \quad  q_r = \frac{1}{2} \sqrt{p}, \quad  s_r =\frac{1}{2} \sqrt{p}
\\ F(q^*, s^* ) &=& \exp \left[ -(1-q^*)^2 -(1-s^* )^2 \right]
\label{Eqn:SelfSimilarity}
\end{eqnarray}
\end{subequations}
%
Thus, knowing $J(p,q,s)$  in a plane $p=\text{const} \le p_{\rm max}$
allows to extrapolate all response functions $J$
for $p > p_{\rm max}$ via equation \eqref{Eqn:Toy:Response}.

\subsection{Support vector regression}
\label{Sec:SVR}

%
In this study,
the analytical response formula from sparse data point
is obtained by support vector regression (SVR) \citep{Cortes1995ml, Drucker1997anips}.
SVR belongs to the family of supervised-learning algorithms
that trains from $M$ observations to find a mapping
between $N$ features or inputs
$\bm{x}_m=[x^1_m,x^2_m,\ldots,x^N_m]$, and the corresponding response $y_m$, $m=1,
\ldots, M$. In the application presented in section \ref{Sec:Results}, the features are the
wavelength $\lambda^+$, period $T^+$, and amplitude $A^+$ and the output is the
relative drag reduction $\Delta c_d$.

Following good practices of machine learning \citep{Burkov2019book},
the inputs for the response formula are  centered features which are normalized to unit variance.
This normalization gives every feature a similar weight in interpolation.
In this study,
the normalization is particularly important
as the ranges of investigated wavelengths and periods
differ by more than one order of magnitude.

SVR yields a regression model $\hat{J}(\bm{x})$
smoothly interpolating from data points $(\bm{x}_m,y_m)$, $m=1,\ldots,M$,
employing a Gaussian Kernel $K \left( \bm{x},\bm{x}_m \right)$
and optimized weights  $\omega_m$:
%
\begin{equation}
  \hat{J}(\bm{x}) = \mu + \sum\limits_{m=1}^M\omega_m K (\bm{x},\bm{x}_m)
  = \mu + \bm{\omega}^\mathrm{T} \bm{K} (\bm{x} ).
\label{Eqn:ReponseModel}
\end{equation}
%
Here, $\mu$ is a constant to which $\hat{J}$ converges far away from the data points,
$\bm{\omega}^\mathrm{T}=[\omega_1,\omega_2,\ldots,\omega_M]$ denotes the weight vector,
and
$\bm{K}^\mathrm{T}=[K(\bm{x},\bm{x}_1),K(\bm{x},\bm{x}_2),\ldots,K(\bm{x},\bm{x}_M)]$
comprises the Gaussian Kernel functions.

Calibrating the response model \eqref {Eqn:ReponseModel}
for $\hat{J} ( \bm{x}_m) = y_m$, $m=1,\ldots,M$
leads to $m$ linear equations for $m$ weights $\omega_m$.
Under generic conditions,
such a linear system can be solved
and the formula will exactly reproduce the input data.
Yet, this vanishing in-sample error
may come at the price of overfitting.
Noise may be incorporated as data feature,
thus leading to an unphysical model complexity.
The over-fitted model may amplify noise outside the training data,
implying a large generalization error or, equivalently, a large out-of-sample error.

To account for noise and new data points,
an error of $\varepsilon$ is tolerated,
i.e.,  a prediction $|\hat{J}(\bm{x}_m) - y_m| < \varepsilon$ is accepted.
The generalization error of the formula
is reduced by avoiding unnecessary complexity,
e.g., by replacing two Kernels of very close collocation points with a single one.
Complexity is characterized and penalized by the vector norm $\Vert \bm{w} \Vert^2$.
This leads to the regularized optimization problem
\begin{equation}
\begin{aligned}
  \min  \quad & \frac{1}{2} \left\Vert \bm{\omega} \right\Vert^2 \\
  \text{subject to} \quad & \left \vert y_m - \bm{\omega}^\mathrm{T}\bm{K} - \mu \right \vert \leq \varepsilon.\\
\end{aligned}
\label{Eq:SVR:Opt1}
\end{equation}

However, weights $\bm{\omega}$  which satisfy the $\varepsilon$-constraint
at all points $(\bm{x}_m,y_m) $ might not exist,
particularly for the validation data.
This constraint is relaxed by introducing so-called
\emph{slack variables} $\xi^+_m$ in case $\hat{J}(\bm{x}_m) - y_m > \varepsilon$
and $\xi^-_m$  if $y_m - \hat{J}(\bm{x}_m)  < \varepsilon$.
The slack variables extend the permissible $\varepsilon$-interval for ${\hat J} ( \bm{x}_m ) $
to $\left [ y_m-\varepsilon-\xi_m^-, y_m+\varepsilon + \xi_m^+ \right]$.
Now, the relaxed regularized optimization problem becomes
%
\begin{equation}
  \begin{aligned}
    \min  \quad & \frac{1}{2}\Vert \bm{\omega} \Vert^2 + C \frac{1}{M}\sum\limits_{m=1}^M(\xi_m^+ +\xi_m^-)\\
    \text{subject to} \quad & y_m - \bm{\omega}^\mathrm{T}\bm{K} - \mu \leq \varepsilon + \xi_m^+\\
                            & \bm{\omega}^\mathrm{T}\bm{K} + \mu - y_m \leq \varepsilon + \xi_m^-\\
                            & \xi_m^+, \quad \xi_m^-\geq 0.\\
  \end{aligned}
  \label{Eq:SVR:Opt2}
\end{equation}
%
The tradeoff between model complexity
and errors beyond the $\varepsilon$ limit
is controlled by the penalty parameter $C$.
The extreme choice $C=0$ leads to unpenalized, arbitrarily large slack variables
and minimal complexity since the minimization solely focuses on
$\frac{1}{2}\Vert \bm{\omega} \Vert^2$. 
In other words, $\hat J \equiv \mu$ is a  constant function.
For sufficiently large $C$, the accuracy of the response model is optimized
tolerating maximum complexity.
The value of $\varepsilon$ is set to the data noise level, if available. Note
that too large a value will decrease the prediction accuracy.

The interpolation is performed with the radial basis function
\begin{equation}
  K(\bm{x},\bm{x'}) = \exp{\left(-\frac{|\bm{x}-\bm{x'}|^2}{\sigma^2}\right)}.
\end{equation}
The reader can refer to \citep{Forrester2009pas}
for more details about the SVR formulation and the solution of the constrained
optimization problem (\ref{Eq:SVR:Opt2}).

\subsection{Data-driven response surface}
\label{Sec:Procedure}
The analytical example preludes our data-driven approach
for the actuated turbulent boundary layer.
The approach consists of the following steps:
\begin{description}
  \item[Step 1 ] We consider $M$ computed response function values $J_m$
    for parameter points $\bm{p}_m$, $m=1,\ldots,M$,
    covering well the parameter range of interest $\Omega$.
    Each parameter may need to be centered and scaled to unit variance
    for the regression problem.
  \item[Step 2 ] Interpolate all function values in the domain
    with an accurate and smooth machine learning regression.
  \item[Step 3 ] Apply a gradient search technique for several initial conditions.
    If  the  corresponding steepest ascent curves converge to a point inside the domain,
    the purpose of a response surface model is served.
  \item[Step 4 ] Identify the ridgeline coordinates ($q_r(p), s_r(p)$) and response $J_r(p)$
    leading out of the domain $\Omega$, and model them using simple functions.
    This simple model can now be used to extrapolate the ridgeline outside $\Omega$
    towards the global response optimum.
    Note the choice of the parametrizing ridgeline parameter (here: $p$)
    is problem-dependent.
  \item[Step 5 ] In some cases, like in the example \eqref{Eqn:Example},
    the response function $J$ exhibits self-similar behavior, and can be expressed
    as a $p$-dependent function multiplying a scaled $(q,s)$-dependent function as:
    \begin{equation*}
      \hat{J}(p,q,s) = J_r (p) \> \> F\left ( \frac{q}{q_r(p)}, \frac{s}{s_r(p)} \right),
    \end{equation*}
    where $J_r$ is the ridgeline response, and $F$ is a shape function with the
    maximum $F(1,1)=1$.
    In this case, the shape function and extrapolated ridgeline
    can be used to predict the response $\hat{J}$ to parameter inputs away from the ridgeline.
\end{description}
Note that the parameters $p$, $q$, and $s$ used in this analytical example
correspond to the wavelength $\lambda^+$, period $T^+$, and amplitude $A^+$ for
the boundary layer application.

\section{Results}
\label{Sec:Results}
The previous section discussed the modeling methodology of response functions,
which assumes  the optimal value at the boundary of the explored parameter space.
This section applies the approach to drag reduction for an actuated boundary layer
with spanwise traveling surface waves.
We follow the steps outlined in section \ref{Sec:Methodology}.

\begin{figure}
	\begin{center}
    \includegraphics[width=0.9\linewidth]{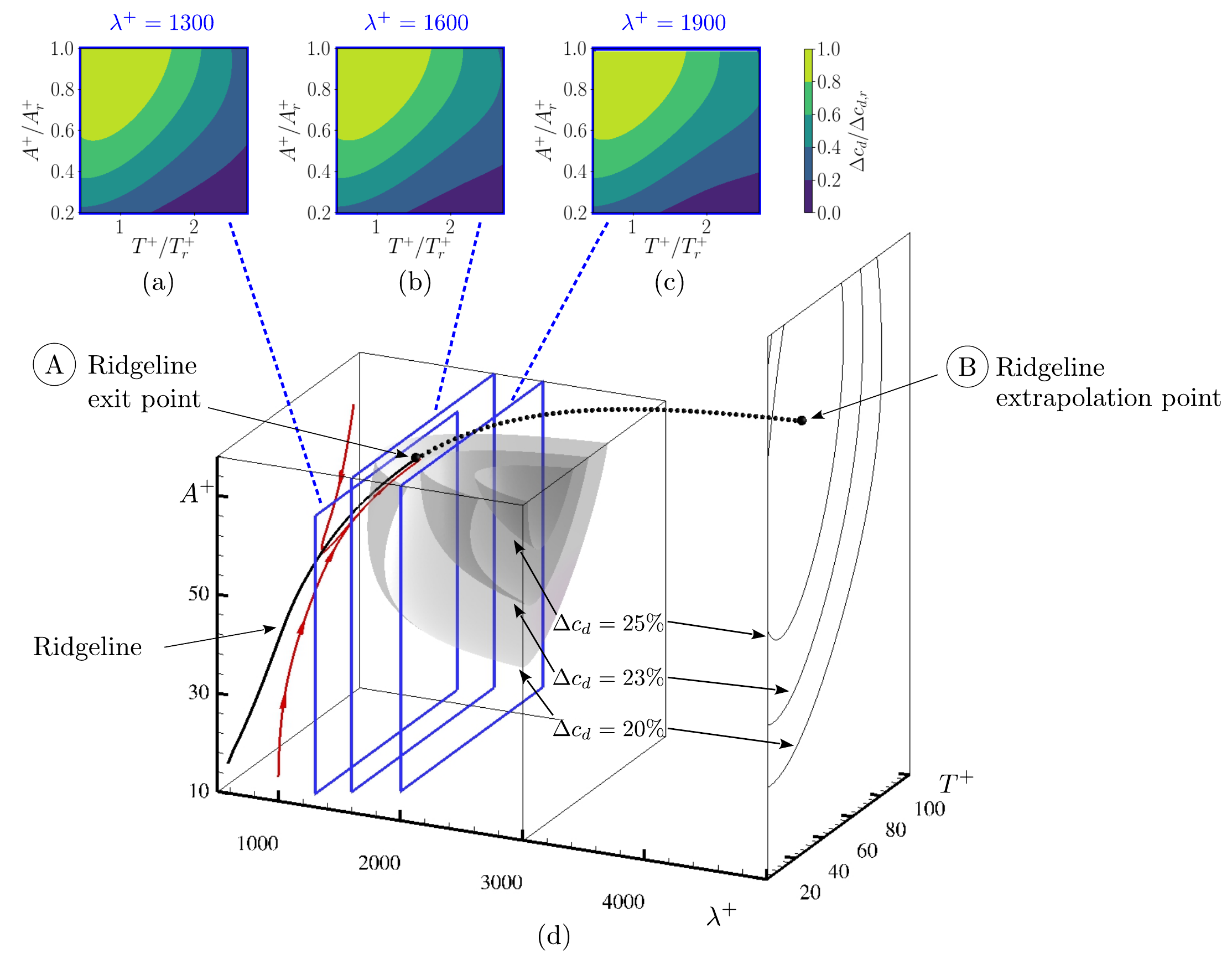}
	\end{center}
  \caption{
    The drag reduction model $\widehat{\Delta c}_D (\lambda^+,T^+,A^+)$.
    The gray surfaces represent three  drag reduction levels: 20\%, 23\%, and 25\%.
    The ridgeline (black) is displayed in the interpolated (solid line)
    and the extrapolated (dotted line) regimes.
    The ridgeline leaves the investigated domain at point $A$
    and predicts the optimal drag reduction for $\lambda^+ \le 5000$ at point
    $B$ ($\lambda^+=5000$, $T^+=44$, and $A^+=99$).
    The red curves denote lines of steepest ascent seeded at various domain locations.
    The contour distributions at the top (($a$)--($c$))
    represent scaled drag reductions (shape functions)
    corresponding to the blue-framed $T^+$--$A^+$ rectangles
    in the three-dimensional plot. }
	\label{Fig:ML:Streamlines}
\end{figure}
%
The process begins with the interpolation of the sparse parameter space using
support vector regression (SVR).
As presented in section \ref{Sec:ComputationalSetup},
the investigated parameter space spanned by \LP{}, \TP{}, and \AP{} is large,
and a dense coverage is beyond the reach of feasibility.
The SVR algorithm is chosen for its prediction accuracy and its
smooth response distribution (see appendix \ref{Sec:MLModel} for details).
The algorithm is trained on a subset of \SI{80}{\percent} of the dataset,
whilst the remaining \SI{20}{\percent} is used for testing the prediction performance.
This separation of training and testing data reduces the risk of overfitting.
The algorithm hyperparameters are tuned using 3-fold cross-validation.
%
%
In this study, the SVR model yields $R^2=0.93$, the definition of which is given
in appendix \ref{Sec:MLModel}.
This value indicates excellent prediction accuracy.
Using the SVR model, we interpolate the parameter space
with drag coefficient predictions.

With the parameter space densely populated, it is now possible to compute and to
visualize the streamlines of the gradient field and the ridgeline.
It is shown in figure \ref{Fig:ML:Streamlines} (d) that the streamlines
(red) and the ridgeline (solid black line) terminate at the domain
boundary in the $T^+$-$\lambda^+$ plane at the exit point $A$ ($\lambda^+=1875$, $T^+=44$, and $A^+=78$).
This indicates that the optimal drag reduction lies outside the current range.
Along the ridgeline, the relative drag reduction increases from $\Delta
c_d=\SI{7.0}{\percent}$ at $\lambda^+=500$ to $\Delta c_d=\SI{22.5}{\percent}$
at the ridgeline exit point $A$.

\begin{figure}
	\begin{center}
	 \includegraphics[width=0.8\linewidth]{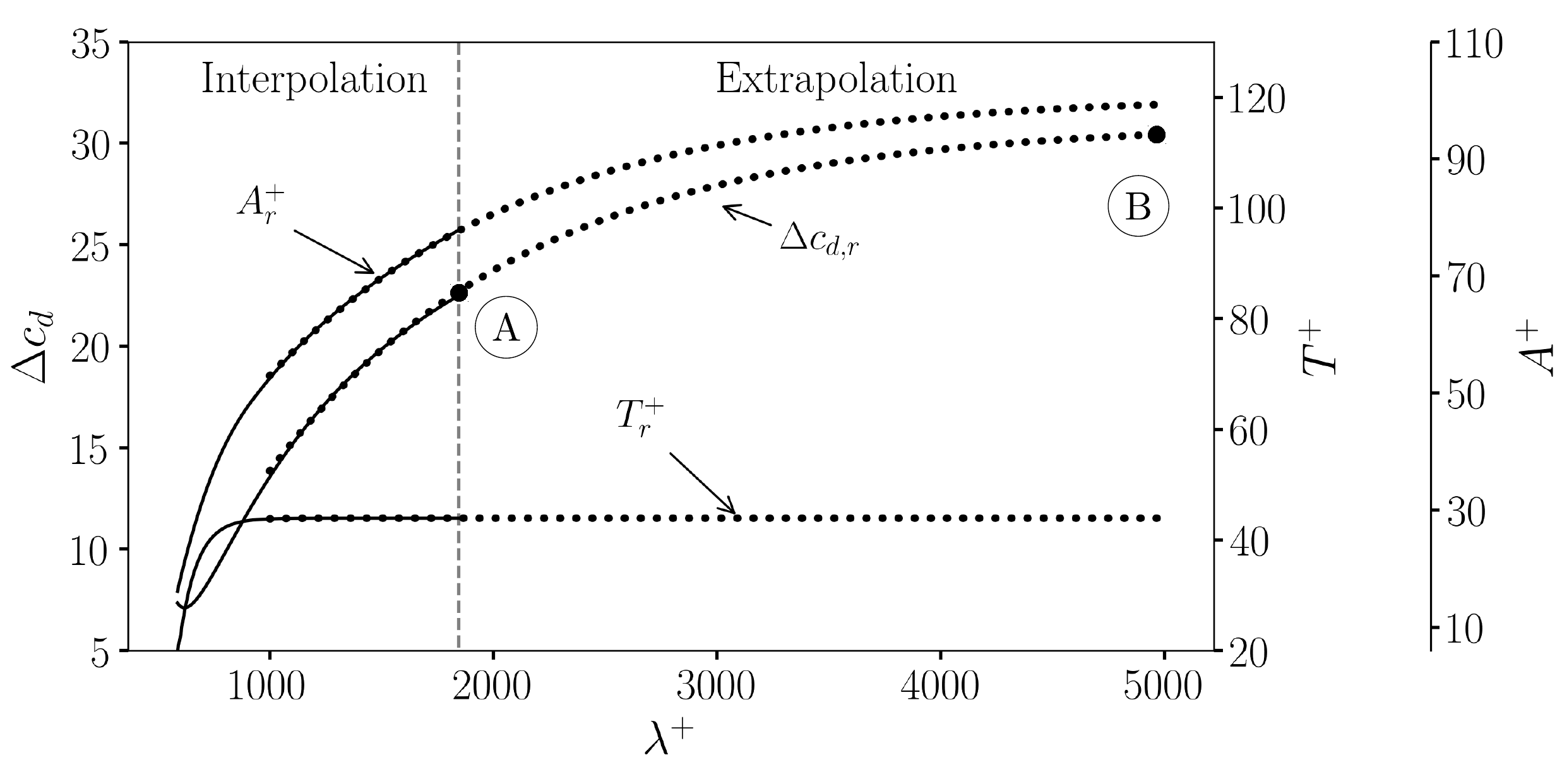}
	\end{center}
	\caption{
    Projection of the ridgeline onto the $\lambda^+$-$T^+$ and $\lambda^+$-$A^+$ plane,
    as well as the drag reduction along the ridge as function of $\lambda^+$.
    The solid lines are interpolated with SVR, whereas the dotted lines are obtained
    by equations (\ref{Eq:PM:MLSACoordinates}) (for
    $T_r^+$ and $A_r^+$), and (\ref{Eq:Tomiyama:LinearFit}) (for $\Delta
    c_{d,r}$). Points $A$ and $B$
    are the same as those in figure \ref{Fig:ML:Streamlines}.
    The vertical grey dashed line separates the interpolation and extrapolation
    regions.
  }
	\label{Fig:PM:Extrapolation}
\end{figure}

\Cref{Fig:PM:Extrapolation} shows the projection of the ridgeline onto the \LP{}--\TP{} and \LP{}--\AP{} planes.
Starting at $\lambda^+\approx1000$, $T_r^+$ asymptotes rapidly toward 44.
In other words, the optimum wave period remains constant at $T^+=44$, even with increasing
wavelength and amplitude. Similarly, $A_r^+$ shows an
asymptotic behavior with higher $\lambda^+$, albeit at a
slower rate.
%

This asymptotic behavior of the ridgeline starting at $\lambda^+=1000$ is easily modeled as
%
\begin{align}
T_r^+ & = 44 - 46721 \exp(-0.0128 \lambda^+) \nonumber \\
A_r^+ & = 100 - 113 \exp(-0.0009 \lambda^+)\>.
\label{Eq:PM:MLSACoordinates}
\end{align}
%
The fitted curves are presented in \cref{Fig:PM:Extrapolation} by dotted lines
and show good agreement with the reference lines over the common range
($1000\leq\lambda^+\leq 1875$).

\begin{figure}
	\begin{center}
	 \includegraphics[width=0.48\linewidth]{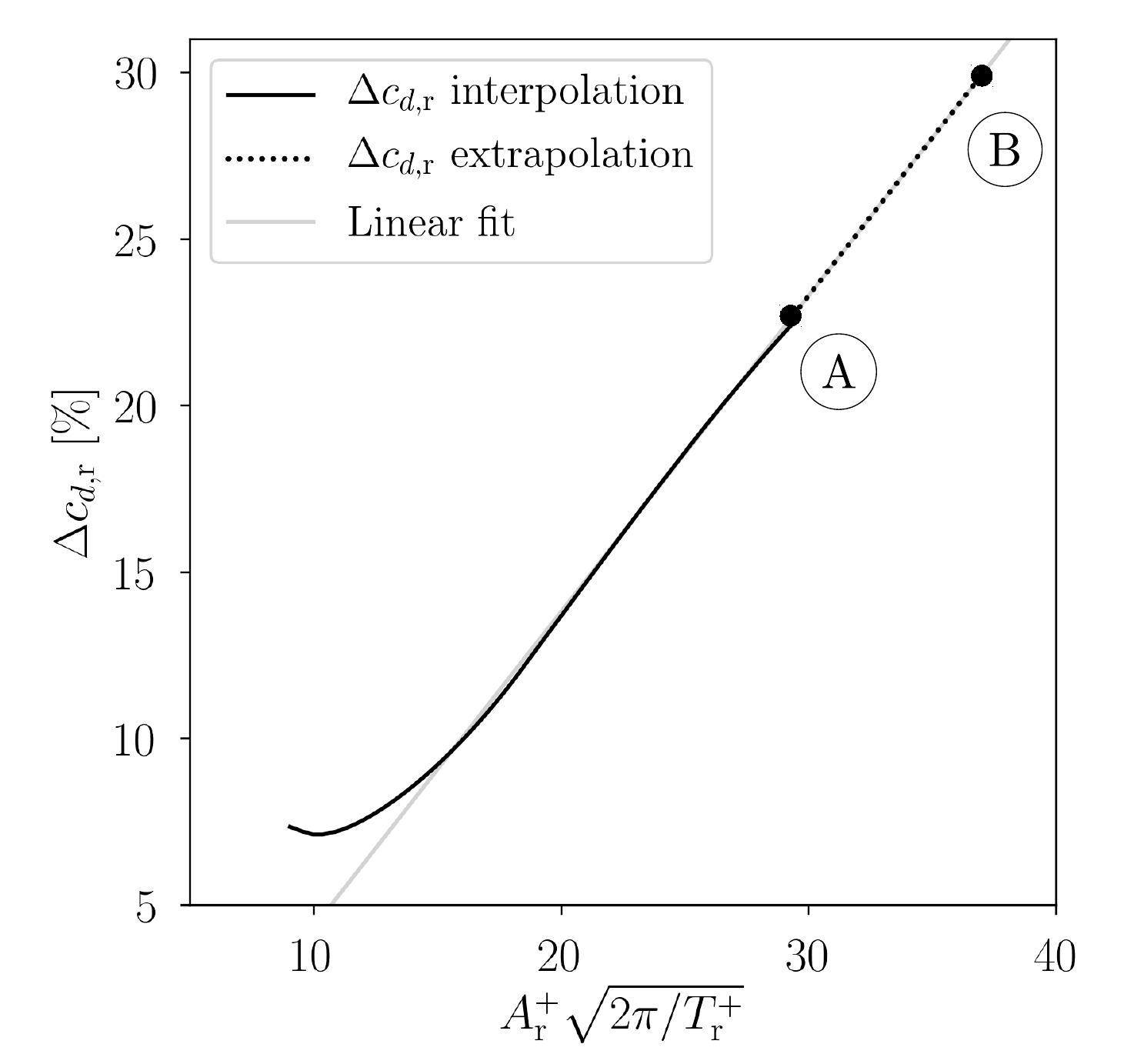}
	\end{center}
	\caption{
    Drag reduction along the ridgeline as function of the Tomiyama and Fukagata scaling. The
    figure shows a linear behavior starting at $\lambda^+\approx1000$.
    The solid line is obtained from data interpolated with SVR. The dotted line is obtained
    with equation (\ref{Eq:Tomiyama:LinearFit}).
    Points $A$ and $B$ are the same as those in figure \ref{Fig:ML:Streamlines}.
  }
	\label{Fig:PM:Tomiyama}
\end{figure}

Having established the $T_r^+$ and $A_r^+$ sole dependence on \LP{} along the ridgeline, we turn our attention to drag reduction.
Similarly to \TP{} and \AP{}, $J_r$ also shows the sole dependence on \LP{} or equivalently on $T_r^+$ and $A_r^+$.
This is best expressed using a scaling proposed by
\cite{tomiyama_direct_2013}, defined as $A_r^+\sqrt{2\pi/T_r^+}$,
which is the product of the velocity amplitude of the
actuation $2\pi A_r^+ / T_r^+$ and the thickness of the Stokes layer
$\sqrt{T_r^+ / (2\pi)}$ along the ridge. Note that this scaling is originally defined for the skin-friction
coefficient, but for the considered cases, the amount of added wetted surface is
negligible and the scaling holds \citep{Albers2019Arxiv}.
The evolution of $\Delta c_{d,r}$ towards a linear behavior is illustrated in \cref{Fig:PM:Tomiyama}.
As the figure shows, the drag reduction along the ridge starts exhibiting
linearity around $\lambda^+\approx1000$ corresponding approximately to
$A^+\sqrt{2\pi/T^+}=19$.
It is worth to note that this almost perfect linear Tomiyama and Fukagata scaling only holds along the ridgeline.
Away from the ridgeline, the scaling shows scatter.
%
%
It is now straightforward to model the
relative drag reduction in the linear range, i.e., $\lambda^+\geq 1000$, as
%
\begin{equation}
\widehat{\Delta c}_{d,r} = 0.95A_r^+\sqrt{\frac{2\pi}{T_r^+}} -5.16.
\label{Eq:Tomiyama:LinearFit}
\end{equation}
%
This linear fit is shown with a dotted line in figure \ref{Fig:PM:Tomiyama} as function of the
Tomiyama and Fukagata scaling, and in figure \ref{Fig:PM:Extrapolation} as a function of
$\lambda^+$.
Note that we assume that the linear behavior continues for a finite range beyond
$\lambda^+=1875$.

Based on the optimal drag reduction behavior being only dependent on $\lambda^+$, which is
consistent with a self-similar behavior, we assume a response of the form
%
\begin{equation}
  \hat{J} =  \widehat{\Delta c}_d \left( \lambda^+, T^+, A^+ \right)=J_r(\lambda^+)\cdot F(T^*,A^*),
  \label{Eq:PM:VariablesSep}
\end{equation}
%
where $\hat{J}_r=\widehat{\Delta c}_{d,r}$ is the constant-linear model from substituting
\cref{Eq:PM:MLSACoordinates} into
\cref{Eq:Tomiyama:LinearFit}, and $T^*$ and $A^*$ are properly scaled actuation parameters.
The natural scaling choice is the maximum relative drag reduction along the ridgeline,
which yields
%
\begin{equation}
  F\left( \frac{T^+}{T_r^+}, \frac{A^+}{A_r^+} \right) = \frac{J(\lambda^+,T^+,A^+)}{J_r(\lambda^+)}.
  \label{Eq:PM:SelfSim}
\end{equation}
%
Hence, self-similarity is validated when $F$ becomes independent of $\lambda^+$.
This is confirmed in figures \ref{Fig:ML:Streamlines} (a), (b), and (c), where the $F$
distributions collapse starting at $\lambda^+\approx 1000$.

Note that the preceding analysis did not only examine the  sensitivities of the flow response and its self-similar behavior,
but also yielded a simple powerful model of the relative drag reductions.

This self-similar drag reduction model for $\lambda^+\geq 1000$  proceeds as follows:
\begin{itemize}
	\item For given actuation setting $\lambda^+$, $T^+$ and $A^+$, compute $T_r$ and $A_r$ from
	 \cref{Eq:PM:MLSACoordinates}.
	\item Determine the drag reduction along the ridgeline $J_r=\Delta c_{d,r}(\lambda^+)$ using \cref{Eq:Tomiyama:LinearFit}.
	\item Read $F(\frac{T^+}{T_r^+}, \frac{A^+}{A_r^+} )$ from the distributions in \cref{Fig:ML:Streamlines}.
	\item Deduce the relative drag reduction from $J=\Delta c_d=J_r\cdot F$.
\end{itemize}

In the interpolation regime, and for $1000\leq\lambda^+\leq3000$, this simple
model has a coefficient of determination of $R^2=0.92$, which is very close to
that of the SVR model.
In the extrapolation regime, the model is validated with two points at
$\lambda^+=5000$, which is well beyond the training range. The first validation point B is situated on the ridgeline (cf. figure
\ref{Fig:ML:Streamlines}), whereas the second point $B'$ is off the ridgeline at
coordinates $\lambda^+=5000$, $T^+=44$, and $A^+=92$. For these two operating
conditions, the relative drag
reductions predicted by the model are $\widehat{\Delta
c}_d=\SI{30.45}{\percent}$ and $\widehat{\Delta c}_d=\SI{30.23}{\percent}$,
which compare favorably with those of the reference LES data of $\Delta
c_d=\SI{31.09}{\percent}$ and  $\Delta
c_d=\SI{30.03}{\percent}$.
These predictions yield relative errors of \SI{2.1}{\percent} and
\SI{0.7}{\percent} for $B$ and $B'$, respectively.

%

The prediction accuracy for the extrapolation at $\lambda^+=5000$,
i.e., \SI{67}{\percent}  beyond the maximal investigated values $\lambda^+=3000$
is impressive.
Yet, the  model (\ref{Eq:PM:VariablesSep}) should not be assumed to hold
at much larger wavelengths ($\lambda^+\to\infty$).
In this limit, the actuation approaches that of a flat plate moving up and down without height variations
in the spanwise direction.
In this scenario, the boundary layer remains
unchanged and no drag reductions can be expected.

\section{Conclusions}
\label{Sec:Conclusions}
We target improved drag reduction  of an actuated turbulent boundary layer with
spanwise traveling surface waves at $Re_{\theta}=1000$.
71 large-eddy simulations are used to determine a machine learned model
to predict drag reduction as a function of the actuation parameters: amplitude,
period, and spanwise wavelength.
The first
enabler for this formula is the support vector regression (SVR) for
smooth interpolation.  For this dataset, SVR is found to be distinctly superior to many other
common regression solvers.  The second enabler is a ridgeline pointing outside the
computed domain indicating further drag reduction potential at unexplored higher
wavelengths.  This ridgeline is then modeled and used for extrapolation.  The
results indicate a potential around \SI{31}{\percent} drag reduction with increasing wavelength,
which is denoted as point B in figure \ref{Fig:ML:Streamlines}. This result is
confirmed by an additional LES.
The corresponding period seems to asymptote against 44
plus units while the amplitude slowly increases with wavenumber.  The ridgeline
parameters are consistent with the Tomiyama and Fukagata scaling.  More precisely, at wavelengths
above 1000 plus units within the analyzed range, the drag reduction linearly increases with the Tomiyama
and Fukagata parameter on the ridgeline.

Surprisingly, the drag reduction formula exhibits a self-similar behavior
starting at $\lambda^+\approx 1000$.  As such, drag reduction can be expressed as the
product of a factor depending only on the wavelength and a shape factor
depending on amplitude and period normalized with their ridgeline values.  The
ridgeline parameters and the drag reduction values in a plane with constant
wavelength allow to extrapolate drag values for amplitudes and periods
for wavelengths above 1000 plus units.  The self-similar drag
reduction formula beautifully parameterizes all investigated simulations and
allow to predict further performance potential at unexplored larger
wavelengths.

The proposed machine learning method
for the drag reduction formula can easily be applied
to other performance parametrics from sparse data.
The strategy is
(1) to interpolate the sparse parameter space
using an accurate machine learning algorithm;
(2) to compute several steepest ascent lines and ridgelines;
(3) to search for the global optimum inside the domain;
(4) if the steepest ascent lines terminate at the boundary
to extrapolate the ridgeline out of the domain;
(5) to test for self-similarity based on this ridgeline.
Self-similarity opens the possibility to extrapolate the performance
away from the ridgeline.

The drag reduction formula may guide future simulations
in search of larger drag reduction.
In addition, the observed self-similarity
guides and constrains future physics-based models.
The authors actively explore these avenues.

\section*{Acknowledgements}

  The research was funded by the Deutsche Forschungsgemeinschaft (DFG)
  in the framework of the research projects SE 2504/2-1, SCHR 309/52 and SCHR
  309/68. The authors gratefully acknowledge the Gauss Centre for
  Supercomputing e.V. (www.gauss-centre.eu) for funding this project
  by providing computing time on the GCS Supercomputers Hazelhen at
  HLRS Stuttgart and JURECA at J\"ulich Supercomputing Centre (JSC).
  BRN acknowledges support from the French National Research Agency (ANR)
  under grant ANR-17-ASTR-0022 (FlowCon).

\begin{appendix}

\section{Machine learning regression model}
\label{Sec:MLModel}

Drag reduction modeling as a function of the actuation
parameters for the actuated boundary layer is a challenging problem.
The complexity of the response topology led to the utilization of machine
learning (ML)  approaches.
For this application, ML is used to model the drag reduction $\Delta c_d$ under
varying actuation conditions ($\lambda^+$, $T^+$ and $A^+$).
ML algorithms are evaluated based on their prediction accuracy, given by the coefficient of determination $R^2$,
defined as
%
\begin{equation}
R^2 = 1 - \frac{\sum_i^N (\Delta c_{d,i}-\widehat{\Delta c}_{d,i})^2}{\sum_i^N (\Delta c_{d,i}-\overline{\Delta c}_d)^2},
\end{equation}
%
where $\Delta c_{d,i}$ are the reference computed data points, $\widehat{\Delta c}_{d,i}$ are the predicted
ones, $\overline{\Delta c}_d$ is the mean of $\Delta c_{d,i}$, and $N$ is the number of samples
in the test set. A value of $R^2=1$ denotes perfect prediction score.
Besides accuracy, the model smoothness is the second criterion for the model selection.
The ML algorithm smoothness was quantified with the total variation (TV) defined as
%
\begin{equation}
  \begin{split}
  y&=\sum\limits_{i=1}^I\sum\limits_{j=1}^J\sum\limits_{k=1}^K \biggl\{
      (\widehat{\Delta c}_{d,i,j,k} - \widehat{\Delta c}_{d,i-1,j,k})^2 \\
   &\qquad\quad + (\widehat{\Delta c}_{d,i,j,k} - \widehat{\Delta c}_{d,i,j-1,k})^2 +
      (\widehat{\Delta c}_{d,i,j,k} - \widehat{\Delta c}_{d,i,j,k-1})^2
    \biggr\}^{\!1/2},
  \end{split}
  \label{Eq:TotalVariation}
\end{equation}
%
where $I$, $J$, and $K$ are the number of discretized points in the $\lambda^+$,
$T^+$, and $A^+$ directions.
A smooth response is indicated by a lower TV value.
Three machine learning algorithms were benchmarked: the $k$-nearest neighbors
(kNN), random forest (RF), and support vector regression (SVR). The
hyperparameters of each algorithm were optimized using cross-validation,
yielding 5 neighbors for kNN and 300 trees for RF. Radial basis functions are
used for SVR.
The $R^2$ and TV
values for the three algorithms are summarized in \cref{Tab:R2-Smooth}.
%
\begin{table}
  \centering
  \caption{Comparison of the prediction accuracy ($R^2$) and smoothness ($TV$)
    of the three tested machine learning algorithms.
  }
  \begin{tabular}{C{2cm}C{2cm}C{2cm}}
    \hline
    \hline
    Algorithm& $R^2$ & $TV$ \\
    \hline
    kNN & 0.76 & $1.62.10^6$ \\
    RF  & 0.97 & $1.60.10^6$\\
    SVR & 0.93 & $1.31.10^6$\\
    \hline
    \hline
  \end{tabular}
  \label{Tab:R2-Smooth}
\end{table}
%
Based on the results, SVR offers the best compromise between smoothness and
accuracy; it is smoother than RF and more accurate than kNN. Therefore, it is selected
for this study.


\section{Operating conditions of the LES simulations}
\label{Sec:OCLES}
\begin{center}
\tablefirsthead{%
  \hline
  \hline
  $N$ & $L_z^+$ & $\lambda^+$ & $T^+$ & $A^+$ & $\Delta c_d~[ \% ]$ & $\Delta c_f~[ \% ]$  & $\Delta A_\mathrm{surf}~[ \% ]$ \\
  \hline
}
\tablehead{%
  \hline
  \hline
  $N$ & $L_z^+$ & $\lambda^+$ & $T^+$ & $A^+$ & $\Delta c_d~[ \% ]$ & $\Delta c_f~[ \% ]$  & $\Delta A_\mathrm{surf}~[ \% ]$ \\
  \hline}
\tablelasttail{\hline\hline}
\topcaption{
  Actuation parameters of the turbulent boundary layer simulations, where each
  setup is denoted by a case number $N$. The quantity $\lambda^+$ is the
  spanwise wavelength of the traveling wave, $T^+$ is the period, and $A^+$ is
  the amplitude, all given in inner units, i.e., non-dimensionalized by the
  kinematic viscosity $\nu$ and the friction velocity $u_\tau$. Each block
  includes setups with varying period and amplitude for a constant wavelength.
  The list includes the values of the averaged relative drag reduction $\Delta
  c_d$, the averaged relative skin friction reduction  $\Delta c_f$, and the
  relative increase of the wetted surface $\Delta A_\mathrm{surf}$.
}

\begin{supertabular}{
    C{10mm}
    C{10mm}
    C{10mm}
    C{10mm}
    C{10mm}
    C{10mm}
    C{20mm}
    C{20mm}
    C{20mm}
  }
1  & 1000 & 500 & 20 & 30 & 0 & 4 & 3.5 \\
2  & 1000 & 500 & 30 & 22 & 9 & 10 & 1.9 \\
3  & 1000 & 500 & 40 & 21 & 8 & 9 & 1.7 \\
4  & 1000 & 500 & 40 & 30 & 8 & 11 & 3.5 \\
5  & 1000 & 500 & 60 & 30 & 5 & 8 & 3.5 \\
6  & 1000 & 500 & 70 & 36 & 3 & 8 & 4.9 \\
7  & 1000 & 500 & 70 & 64 & -10 & 4 & 14.6 \\
8  & 1000 & 500 & 100 & 48 & -3 & 5 & 8.6 \\
9  & 1000 & 1000 & 20 & 10 & 5 & 5 & 0.1 \\
10 & 1000 & 1000 & 20 & 30 & 13 & 13 & 0.9 \\
11 & 1000 & 1000 & 20 & 50 & 0 & 3 & 2.4 \\
12 & 1000 & 1000 & 40 & 10 & 3 & 3 & 0.1 \\
13 & 1000 & 1000 & 40 & 20 & 7 & 8 & 0.4 \\
14 & 1000 & 1000 & 40 & 30 & 12 & 13 & 0.9 \\
15 & 1000 & 1000 & 40 & 40 & 15 & 16 & 1.6 \\
16 & 1000 & 1000 & 40 & 50 & 15 & 17 & 2.4 \\
17 & 1000 & 1000 & 40 & 60 & 13 & 16 & 3.5 \\
18 & 1000 & 1000 & 80 & 10 & 1 & 1 & 0.1 \\
19 & 1000 & 1000 & 80 & 20 & 3 & 4 & 0.4 \\
20 & 1000 & 1000 & 80 & 30 & 6 & 6 & 0.9 \\
21 & 1000 & 1000 & 80 & 40 & 9 & 10 & 1.6 \\
22 & 1000 & 1000 & 80 & 50 & 9 & 11 & 2.4 \\
23 & 1000 & 1000 & 80 & 60 & 9 & 12 & 3.5 \\
24 & 1000 & 1000 & 120 & 10 & 1 & 1 & 0.1 \\
25 & 1000 & 1000 & 120 & 20 & 0 & 1 & 0.4 \\
26 & 1000 & 1000 & 120 & 30 & 3 & 4 & 0.9 \\
27 & 1000 & 1000 & 120 & 40 & 3 & 5 & 1.6 \\
28 & 1000 & 1000 & 120 & 50 & 2 & 5 & 2.4 \\
29 & 1000 & 1000 & 120 & 60 & 2 & 6 & 3.5 \\
\hline
30 & 1200 & 600 & 30 & 44 & 2 & 7 & 5.1 \\
31 & 1200 & 600 & 40 & 59 & -4 & 5 & 8.9 \\
32 & 1200 & 600 & 50 & 36 & 9 & 12 & 3.5 \\
33 & 1200 & 600 & 60 & 21 & 5 & 6 & 1.2 \\
34 & 1200 & 600 & 70 & 29 & 6 & 8 & 2.3 \\
35 & 1200 & 600 & 80 & 66 & -5 & 6 & 11.0 \\
36 & 1200 & 600 & 90 & 51 & -1 & 6 & 6.8 \\
37 & 1200 & 600 & 100 & 14 & 2 & 2 & 0.5 \\
\hline
38 & 1600 & 1600 & 20 & 22 & 11 & 11 & 0.2 \\
39 & 1600 & 1600 & 40 & 34 & 14 & 14 & 0.4 \\
40 & 1600 & 1600 & 40 & 48 & 19 & 19 & 0.9 \\
41 & 1600 & 1600 & 50 & 60 & 19 & 20 & 1.4 \\
42 & 1600 & 1600 & 50 & 73 & 21 & 22 & 2.0 \\
43 & 1600 & 1600 & 60 & 27 & 8 & 8 & 0.3 \\
44 & 1600 & 1600 & 70 & 71 & 17 & 19 & 1.9 \\
45 & 1600 & 1600 & 80 & 17 & 2 & 2 & 0.1 \\
46 & 1600 & 1600 & 90 & 65 & 13 & 14 & 1.6 \\
47 & 1600 & 1600 & 100 & 40 & 8 & 8 & 0.6 \\
\hline
48 & 1800 & 900 & 30 & 49 & 10 & 12 & 2.9 \\
49 & 1800 & 900 & 40 & 63 & 7 & 12 & 4.7 \\
50 & 1800 & 900 & 50 & 22 & 7 & 7 & 0.6 \\
51 & 1800 & 900 & 50 & 44 & 12 & 14 & 2.3 \\
52 & 1800 & 900 & 70 & 28 & 7 & 8 & 0.9 \\
53 & 1800 & 900 & 80 & 17 & 3 & 4 & 0.4 \\
54 & 1800 & 900 & 80 & 60 & 6 & 9 & 4.3 \\
55 & 1800 & 900 & 90 & 39 & 6 & 7 & 1.8 \\
56 & 1800 & 1800 & 30 & 14 & 5 & 5 & 0.1 \\
57 & 1800 & 1800 & 40 & 51 & 19 & 20 & 0.8 \\
58 & 1800 & 1800 & 40 & 70 & 22 & 23 & 1.5 \\
59 & 1800 & 1800 & 50 & 59 & 20 & 21 & 1.1 \\
60 & 1800 & 1800 & 60 & 44 & 15 & 15 & 0.6 \\
61 & 1800 & 1800 & 60 & 75 & 21 & 22 & 1.7 \\
62 & 1800 & 1800 & 70 & 29 & 7 & 7 & 0.3 \\
63 & 1800 & 1800 & 80 & 36 & 9 & 9 & 0.4 \\
64 & 1800 & 1800 & 90 & 66 & 13 & 14 & 1.3 \\
65 & 1800 & 1800 & 100 & 21 & 3 & 3 & 0.1 \\
\hline
66 & 3000 & 3000 & 40 & 51 & 21 & 21 & 0.3 \\
67 & 3000 & 3000 & 50 & 78 & 26 & 26 & 0.7 \\
68 & 3000 & 3000 & 60 & 26 & 7 & 7 & 0.1 \\
69 & 3000 & 3000 & 70 & 64 & 19 & 19 & 0.4 \\
70 & 3000 & 3000 & 80 & 11 & 1 & 1 & 0.0 \\
71 & 3000 & 3000 & 90 & 66 & 16 & 16 & 0.5 \\
\hline
$B$ & 5000 & 5000 & 44 & 99 & 31 & 31 & 0.0   \\
$B'$& 5000 & 5000 & 44 & 92 & 30 & 30 & 0.0   \\
\end{supertabular}
\end{center}

\end{appendix}

\bibliographystyle{plainnat}
\bibliography{Main} 

\end{document}